\documentclass[12pt]{article}

\usepackage{xspace}
\usepackage{color}
\usepackage{amssymb}
\usepackage{epsfig}

\newcommand{\ppbar}{p{\bar p}}
\newcommand{\invpb}{\rm pb^{-1}} 
\newcommand{\bfinvpb}{\bf pb^{-1}} 
\newcommand{\roots}{{\sqrt s}}
 
\newcommand{\Et}{\rm E_T}
\newcommand{\Pt}{\rm p_T}
\newcommand{\Ht}{H_T}
\newcommand{\egg}{e\gamma\gamma}
\newcommand{\mugg}{\mu\gamma\gamma}
\newcommand{\eeggmet}{ee\gamma\gamma\met}
\newcommand{\llggmet}{\ell\ell\gamma\gamma\met}
\newcommand{\lgal}{\ell\gamma}
\newcommand{\lgX}{\ell\gamma\plus X}
\newcommand{\ljX}{\ell j\plus X}
\newcommand{\leX}{\ell e\plus X}
\newcommand{\rrr}{\rightarrow}

\newcommand{\emugX}{e\mu\gamma+X}

\newcommand{\gt}{>}

\newcommand{\met}{{\rm\not\!\!E}_{T}}
\newcommand{\metvec}{{\not\!\! \vec{E}_T}}

\newcommand{\lplm}{\ell^+\ell^-}
\newcommand{\lgmet}{\ell\gamma\met}

\newcommand{\eg}{e\gamma}
\newcommand{\mug}{\mu\gamma}

\newcommand{\llg}{\ell\ell\gamma}
\def\ttg{t\bar{t}\gamma}
\newcommand{\lgg}{\ell\gamma\gamma}

\newcommand{\eeg}{ee\gamma}
\newcommand{\mumug}{\mu\mu\gamma}

\newcommand{\Zee}{Z^0 \rightarrow e^+e^-}

\newcommand{\Wg}{W \gamma}
\newcommand{\Zg}{Z^0 \gamma}
\newcommand{\Wgg}{W \gamma\gamma}
\newcommand{\Zgg}{Z^0 \gamma\gamma}
\newcommand{\Z}{Z^0}
\newcommand{\Zgstar}{Z^0\kern -0.25em/\kern -0.15em\gamma^*}
\newcommand{\epem}{e^+e^-}
\newcommand{\runonelumi}{86}
\newcommand{\goes}{\kern -0.18em\rightarrow\kern -0.18em}
\newcommand{\plus}{\kern -0.18em +\kern -0.18em}

\newcommand{\MeV}{\ensuremath{\mathrm{\ Me\kern -0.1em V}}\xspace}
\newcommand{\MeVc}{\ensuremath{\mathrm{\ Me\kern -0.1em V\kern -0.1em 
/\mathit{c}}}\xspace}
\newcommand{\MeVcsq}{\ensuremath{\mathrm{\ Me\kern -0.1em V\kern -0.1em 
/\mathit{c}^2}}\xspace}
\newcommand{\GeV}{\ensuremath{\mathrm{Ge\kern -0.1em V}}\xspace}
\newcommand{\GeVc}{\ensuremath{\mathrm{\ Ge\kern -0.1em V\kern -0.1em 
/\mathit{c}}}\xspace}
\newcommand{\GeVcsq}{\ensuremath{\mathrm{\ Ge\kern -0.1em V\kern -0.1em 
/\mathit{c}^2}}\xspace}
\newcommand{\TeV}{\ensuremath{\mathrm{Te\kern -0.1em V}}\xspace}
\newcommand{\bfTeV}{\ensuremath{\bf{Te\kern -0.1em V}}\xspace}
\newcommand{\nsGeV}{\ensuremath{\mathrm{Ge\kern -0.1em V}}\xspace}
\newcommand{\nsGeVc}{\ensuremath{\mathrm{Ge\kern -0.1em V\kern -0.1em 
/\mathit{c}}}\xspace}
\newcommand{\Etgamma}{\ensuremath{\mathrm{E_T^{\gamma}}}}
\newcommand{\Etlepton}{\ensuremath{\mathrm{E_T^{\ell}}}}
\newcommand{\Etelectron}{\ensuremath{\mathrm{E_T^{e}}}}

\input{total.summary}
\begin{document}

\title{\Large \bf 
Search for New Physics in Lepton + Photon + X
 Events with $\bf\lumi$ $\bfinvpb$ of $\bf\ppbar$ Collisions at $\bf\roots$= 1.96 $\bfTeV$
}
\author{\large A.Loginov \thanks{\emph{For the CDF Collaboration}} \bigskip \\
{\it ITEP, Moscow}}

\maketitle

{\large

\begin{center}
{\bf Abstract}\\
\end{center}
\medskip
We present results of a search in $\lumi\pm\dlumi$ $\invpb$ of
$\ppbar$ collisions at 1.96 $\TeV$ for the anomalous production of
events containing a charged lepton ($\ell$, either $e$ or $\mu$) and a
photon ($\gamma$), both with high transverse momentum, accompanied by
additional signatures, X, including missing transverse energy ($\met$)
and additional leptons and photons. We use the same selection criteria
as in the previous CDF Run I search, but with an order-magnitude
larger data set, a higher $\ppbar$ collision energy,
and the CDF II detector. 

\section{Introduction}
\label{introduction}

The signature of high-$E_T$~\cite{EtPt} leptons and photons ($\lgX$)
appears in a variety of physics beyond the Standard Model
(SM)~\cite{SM}, so called New Physics (NP) scenarios, such as
Technicolor, Gauge-Mediated Supersymmetry, LED.


While it is good to be guided by theory, one should also remain open
to the unexpected. Therefore we use the technique of a Signature-Based
Search, and look for significant deviations from the
SM~\cite{Toback_all,Jeff_all} in
$\lgX$ events, which include the production of fundamental particles,
such as $\gamma$, $Z^0$, $W^{\pm}$.

In Run I, in a sample of $\runonelumi$ $\invpb$ of $\ppbar$ collisions at an
energy of 1.80 $\TeV$, the CDF experiment observed a single clean
event consistent with having a pair of high-$\Et$ electrons, two
high-$\Et$ photons, and large $\met$~\cite{Toback_all}. A subsequent
search for `cousins' of the $\eeggmet$ signature in the inclusive
signature $\lgX$ found 16 events with a SM
expectation of 7.6 $\pm$ 0.7 events, corresponding in likelihood to a
2.7 $\sigma$ effect~\cite{Jeff_all}.

In this proceedings we report the preliminary results of extending the
Run II $\lgX$ search~\cite{Loginov_all} to the full data set taken
during the period March 2002 through February, 2006, an exposure of
$\lumi\pm\dlumi$ $\invpb$. We also present the details of the results
for the $\emugX$ signature.

\section{The CDF II Detector}
\label{detector}

The CDF II detector is a cylindrically symmetric spectrometer designed
to study $\ppbar$ collisions at the Fermilab Tevatron based on the
same solenoidal magnet and central calorimeters as the CDF I
detector~\cite{CDFI} from which it was upgraded. Because the analysis
described here is intended to repeat the Run I search as closely as
possible, we note especially the differences from the CDF I detector
relevant to the detection of leptons, photons, and $\met$. The
tracking systems used to measure the momenta of charged particles have
been replaced with a central outer tracker (COT) with smaller drift
cells~\cite{COT}, and an enhanced system of silicon strip
detectors~\cite{SVX}. The calorimeters in the regions~\cite{CDF_coo}
with pseudorapidity $|\eta| \gt 1$ have been replaced with a more
compact scintillator-based design, retaining the projective
geometry~\cite{cal_upgrade}.  The coverage in $\varphi$ of the CMP and
CMX muon systems~\cite{muon_systems} has been extended; the CMU system
is unchanged.

\section{Selection of $\lgX$ Events}
\label{selection}

The identification of leptons and photons is essentially the same as
in the Run I search~\cite{Jeff_all}, with only minor technical changes. 
The identification criteria for the results
for the full dataset presented here are identical to those used for
the first third, presented in Refs~\cite{Loginov_all}.


Inclusive $\lgal$ events are selected by requiring a central
($|\eta|\lesssim 1.0$) $\gamma$ candidate with $\Etgamma>25~\GeV$ and
a central $e$ or $\mu$ with $\Etlepton>25$ $\GeV$.

Additional leptons in the central region are required to have
$\Etlepton>20~\GeV$, and electrons in endplug calorimeters should have
$\Etelectron>15~\GeV$.

Missing transverse energy $\met$ is calculated from the calorimeter
tower energies in the region $|\eta| < 3.6$. Corrections are then made
to the $\met$ for non-uniform calorimeter response~\cite{jet_corr} for
jets with uncorrected $\Et > 15$ $\GeV$ and $\eta < 2.0$, and for
muons with $\Pt > 20$ $\GeV$.

The variable $\Ht$ is defined for each event as the sum of the
transverse energies of the leptons, photons, jets, and $\met$.

\begin{figure}[!t]
\begin{center}
\vspace*{-0.1in}
\includegraphics*[angle =90,width=0.6\textwidth]{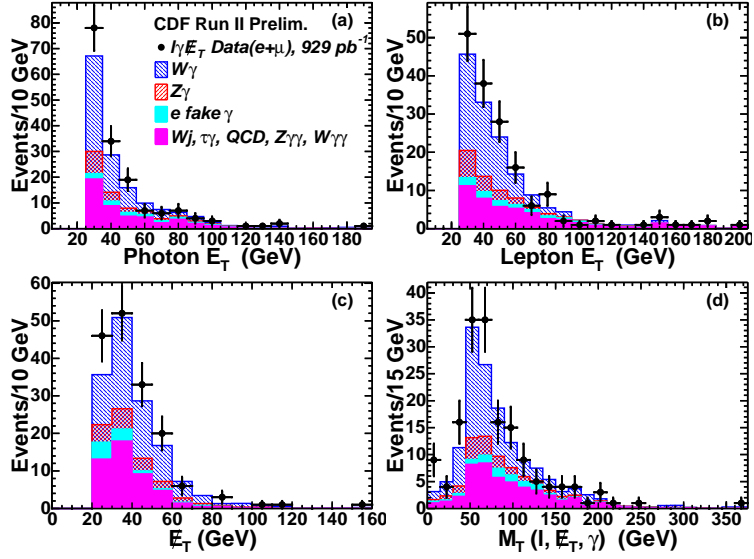}
\end{center}
\caption{ The distributions for events in the $\lgmet$ sample
  (points) in a) the $\Et$ of the photon; b) the $\Et$ of the lepton (e or $\mu$); c)
  the missing transverse energy, $\met$; and d) the transverse mass of
  the $\lgmet$ system.  The histograms show the expected SM
  contributions, including estimated backgrounds from misidentified
  photons and leptons.}
\label{wg_fig1_leptons}
\end{figure}
\begin{figure}[!t]
\begin{center}
\vspace*{-0.1in}
\includegraphics*[angle =90,width=0.6\textwidth]{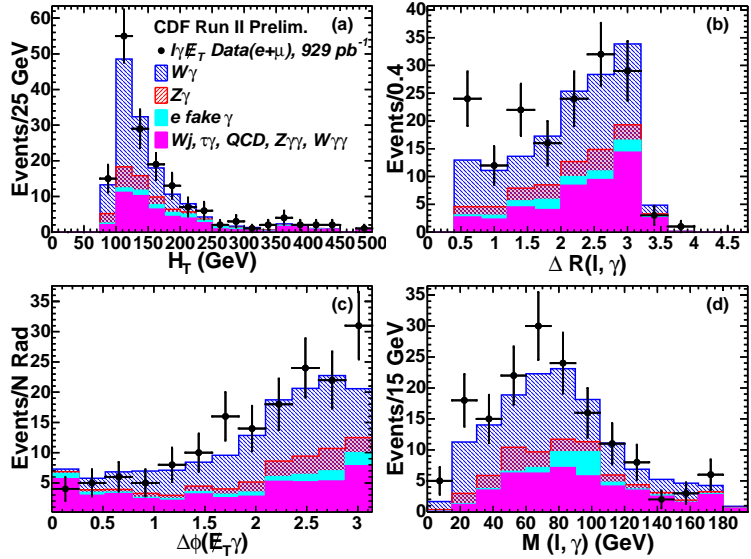}
\end{center}
\caption{ The distributions for events in the $\lgmet$ sample (points)
  in a) $\Ht$, the sum of the transverse energies of the lepton,
  photon, jets and $\met$; b) the distance in $\eta$-$\phi$ space
  between the photon and lepton; c) the angular separation in $\phi$
  between the lepton and the missing transverse energy, $\met$; and d)
  the invariant mass of the $\lg$ system. The histograms show the
  expected SM contributions, including estimated backgrounds from
  misidentified photons and leptons.}
\label{wg_fig2_leptons}
\end{figure}

Because we are looking for processes with small cross sections, and
hence small numbers of measured events, we use larger control samples
to validate our understanding of the detector performance and to
measure efficiencies and backgrounds.

We use $W^\pm$ and $Z^0$ production derived from the same inclusive
lepton datasets as control samples to ensure that the efficiencies for
high-$\Pt$ electrons and muons are well understood. In addition, the
$W^\pm$ samples provide the control samples for the understanding of
$\met$. 
The photon control sample is constructed from $\Zee$ events in which
one of the electrons radiates a high-$\Et$ $\gamma$ such that the
$\eg$ invariant mass is within 10 $\GeV$ of the $Z^0$ mass.


The first search we perform is in the $\lgmet+X$ subsample, defined by
requiring that an event contain $\met> 25~\GeV$ in addition to the
$\gamma$ and ``tight'' lepton. 

The electron and muon kinematic distributions are combined in
Figures~\ref{wg_fig1_leptons} and~\ref{wg_fig2_leptons}. There is very
good agreement with the expected SM shapes~\cite{no_overflows}.


\begin{figure}[!t]
\begin{center}
\vspace*{-0.1in}
\includegraphics*[angle=90,width=0.6\textwidth]{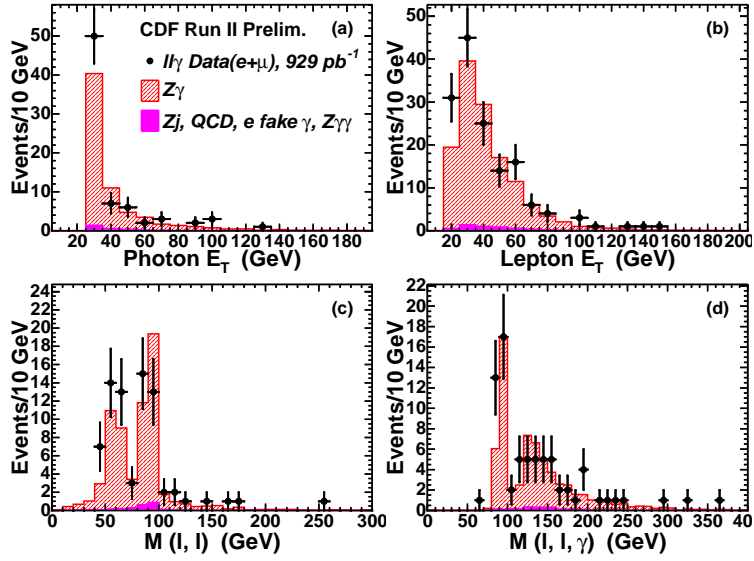}
\end{center}
\caption{ The distributions for events in the $\llg$ sample (points)
in a) the $\Et$ of the photon; b) the $\Et$ of the leptons (two
entries per event); c) the 2-body mass of the dilepton system; and d)
the 3-body mass $m_{\llg}$. The histograms show the expected SM
contributions.}
\label{zg_fig1_leptons}
\end{figure}
\begin{figure}[!t]
\vspace*{-0.1in}
\begin{center}
\hspace*{-0.1in}
\includegraphics*[angle=90,width=0.6\textwidth]{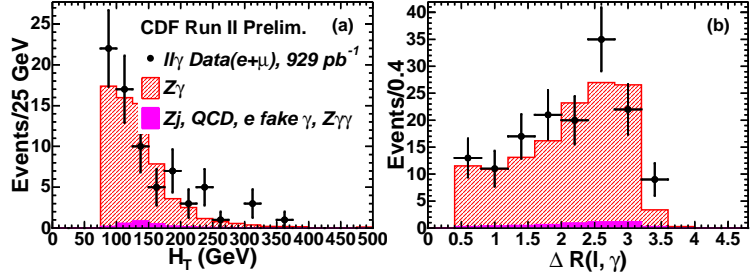}
\end{center}
\caption{ The distributions for events in the $\llg$ sample (points)
in a) $\Ht$, the sum of the transverse energies of the lepton, photon,
jets and $\met$; b) the distance in $\eta$-$\phi$ space between the
photon and each of the two leptons. The histograms show the expected
SM contributions, including estimated backgrounds from misidentified
photons and leptons.}
\label{zg_fig2_leptons}
\end{figure}

A second search, for the $\llg+X$ signature, is constructed by
requiring another $e$ or $\mu$ in addition to the ``tight'' lepton and
the $\gamma$. 
The combined
distributions for electrons and muons are shown in
Figures~\ref{zg_fig1_leptons}
and~\ref{zg_fig2_leptons}~\cite{no_overflows}.

\begin{figure}[!t]
\vspace*{-0.1in}
\begin{center}
\includegraphics*[width=0.22\textwidth, angle=90,clip=]{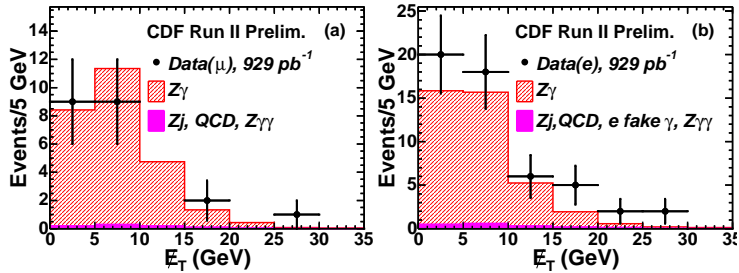}
\end{center}
\caption{The distributions in missing transverse energy $\met$
  observed in the inclusive search for a) $\mumug$ events and
  b) $\eeg$ events. The histograms show the expected SM
  contributions.}
\label{zg_fig3_leptons}
\end{figure}

We do not expect SM events with large $\met$ in the $\llg$ sample; the
Run I $\eeggmet$ event was of special interest in the context of
supersymmetry~\cite{susy} due to the large value of $\met$ (55 $\pm$ 7
$\GeV$). Figure~\ref{zg_fig3_leptons} shows the distributions in
$\met$ for the $\mumug$ and $\eeg$ subsamples of the $\llg$ sample. We
observe 3 $\llg$ events with $\met > 25$ GeV.

\section{Standard Model Expectations}
\label{sm}


The dominant SM source of $\lgal$ events is electroweak $W$ and
$\Zgstar$ production along with a $\gamma$ radiated from one of the
charged particles involved in the process~\cite{CDF_WZgamma}. The
number of such events is estimated using leading-order (LO) event
generators~\cite{MadGraph,Baur,CompHep}. Initial state radiation is
simulated by the {\sc pythia} shower Monte Carlo (MC)
code~\cite{Pythia} tuned to reproduce the underlying event. The
generated particles are then passed through a full detector
simulation, and these events are then reconstructed with the same code
used for the data. 
We have used both {\sc madgraph}~\cite{MadGraph} and {\sc
comphep}\cite{CompHep} to simulate the triboson channels $\Wgg$ and
$Z\gamma\gamma$. A correction for higher-order processes
(K-factor) that depends on both the dilepton mass and photon $\Et$ has
been applied~\cite{Baur_NLO}. to $\Wg$, $\Zg$, $\Wgg$ and $\Zgg$.

High $\Pt$ photons are copiously created from hadron decays in jets
initiated by a scattered quark or gluon. In particular, mesons such as
the $\pi^0$ or $\eta$ decay to photons which may satisfy the photon
selection criteria. The numbers of $\ljX$, $j\rrr\gamma$ events
expected in the $\lgmet$ and $\llg$ samples are determined by
measuring energy in the calorimeter nearby the photon candidate. 

The numbers of $\ljX$, $j\rrr\gamma$ events expected in the
$\lgg$ and $e\mug$ samples are determined by measuring the jet $\Et$
spectrum in $\lgal+$jet, $\ell+$jets and $e\mu+$jet samples,
respectively, and then multiplying by the probability of a jet being
misidentified as a photon, $P^{jet}_{\gamma}(\Et)$, which is measured
in data samples triggered on jets~\cite{CDF_WZgamma}. 

To estimate the numbers of $\leX$, $e\rrr\gamma$ expected events we
measure $P^{e}_{\gamma}$, the probability that an electron undergoes
hard bremsstrahlung and is misidentified as a photon. $P^{e}_{\gamma}$
is measured from the photon control sample. 
We then apply this
misidentification rate to electrons in the $\leX$ samples.



We have estimated the background due to events with jets misidentified
as $\lgmet$ or $\llg$ signatures by studying the total $\Pt$ of tracks
in a cone in $\eta-\varphi$ space of radius $R=0.4$ around the lepton
track~\cite{QCD_background}. 


There is a muon background that we expect escapes the above method: a
low-momentum hadron, not in an energetic jet, decays to a muon in a
configuration that a high-momentum track is reconstructed from the
initial track segment due to the hadron and the secondary track
segment from the muon~\cite{decay_in_flight}. The contribution from
this background is estimated by identifying tracks consistent with a
``kink'' in the COT. 

%
\section{Results}
\label{results}

\begin{table}[!t]
\begin{center}
\caption{A comparison of the numbers of events predicted by the
SM and the observations for the $\lgmet$ signature. The SM predictions
are dominated by $\Wg$ and $\Zg$
production~\cite{MadGraph,Baur,CompHep}. Other contributions come from
$\Wgg$ and $\Zgg$, leptonic $\tau$ decays, and misidentified leptons,
photons, or $\met$.}
\begin{tabular}{|l|c|c||c|}
\hline
\multicolumn{4}{|c|} {
{\bf CDF Run II Preliminary, $\apprluminosity\invpb$}}\\[3pt]
\hline
\multicolumn{4}{|c|} {
Lepton+ Gamma +$\met$ Events}\\
\hline
Standard Model Source & $e\gamma\met$ & $\mu\gamma\met$ & $(e+\mu)\gamma\met$ \\
\hline
$W^{\pm}\gamma$                         & 
$\wanofeglgmet  \pm \totwadsysnofeglgmet$     &
$\wanofmuglgmet \pm \totwadsysnofmuglgmet$    &
$\wanoflglgmet  \pm \totwadsysnoflglgmet$     \\
$\Z/\gamma + \gamma$                    & 
$\zanofeglgmet  \pm \totzadsysnofeglgmet$     &
$\zanofmuglgmet \pm \totzadsysnofmuglgmet$    &
$\zanoflglgmet  \pm \totzadsysnoflglgmet$     \\
$W^{\pm}\gamma\gamma$                   & 
$\waanofeglgmet  \pm \totwaadsysnofeglgmet$   &
$\waanofmuglgmet \pm \totwaadsysnofmuglgmet$  &
$\waanoflglgmet  \pm \totwaadsysnoflglgmet$   \\
$\Z/\gamma + \gamma\gamma$              & 
$\zaanofeglgmet  \pm \totzaadsysnofeglgmet$   &
$\zaanofmuglgmet \pm \totzaadsysnofmuglgmet$  &
$\zaanoflglgmet  \pm \totzaadsysnoflglgmet$   \\
$\ttg$                                  & 
$\ttanofeglgmet  \pm \totttadsysnofeglgmet$   &
$\ttanofmuglgmet \pm \totttadsysnofmuglgmet$  &
$\ttanoflglgmet  \pm \totttadsysnoflglgmet$   \\
\hline
$\Z\goes \epem, e\goes\gamma$           & 
$\efpnofeglgmet \pm \totefpdsysnofeglgmet$    &
--                                       &
$\efpnofeglgmet \pm \totefpdsysnofeglgmet$    \\
Jet faking $\gamma$                     & 
$\nofejglgmet  \pm \totdsysnofejglgmet$       &
$\nofmujglgmet \pm \totdsysnofmujglgmet$      &
$\nofljglgmet  \pm \totdsysnofljglgmet$       \\
$\tau\gamma$ contribution               & 
$\wazataunofeglgmet  \pm \totwazataudsysnofeglgmet $  &
$\wazataunofmuglgmet \pm \totwazataudsysnofmuglgmet$  &
$\wazataunoflglgmet  \pm \totwazataudsysnoflglgmet $  \\
QCD(Jets faking $\ell$ and $\met$)     & 
$\jqcdnofeglgmet  \pm \totjqcddsysnofeglgmet $   &
$\jqcdnofmuglgmet \pm \totjqcddsysnofmuglgmet$   &
$\jqcdnoflglgmet  \pm \totjqcddsysnoflglgmet $   \\
DIF (Decays-In-Flight)                         & 
--   &
$\difnofmuglgmet \pm \totdifdsysnofmuglgmet$   &
$\difnoflglgmet  \pm \totdifdsysnoflglgmet $   \\
\hline
Total                                            & 
$\smnofeglgmet  \pm \totdsysnofeglgmet (tot)$    &
$\smnofmuglgmet \pm \totdsysnofmuglgmet(tot)$    &
$\smnoflglgmet  \pm \totdsysnoflglgmet (tot)$    \\
\hline
Observed in Data                       & 
$\nofeglgmet$                          & 
$\nofmuglgmet$                         & 
$\noflglgmet$                          \\
\hline
\end{tabular}

\label{lgmet.table}
\end{center}
\end{table}

\begin{table}[!t]
\begin{center}
\caption{A comparison of the numbers of events predicted by the
SM and the observations for the $\lgmet$ signature. The SM predictions
are dominated by $\Zg$ production~\cite{MadGraph,Baur,CompHep}. Other
contributions come from $\Zgg$, and misidentified leptons, photons, or
$\met$.}
\begin{tabular}{|l|c|c||c|}
\hline
\multicolumn{4}{|c|} {
{\bf CDF Run II Preliminary, $\apprluminosity\invpb$}}\\[3pt]
\hline
\multicolumn{4}{|c|}{
Multi-Lepton + Photon Predicted Events} \\
\hline
SM Source & $\eeg$ & $\mumug$ & $\llg$ \\
\hline
$\Z\gamma$            		          & 
$\zanofegmultil  \pm \totzadsysnofegmultil$     &
$\zanofmugmultil \pm \totzadsysnofmugmultil$    &
$\zanoflgmultil  \pm \totzadsysnoflgmultil$     \\
$\Z\gamma\gamma$ 	     	& 
$\zaanofegmultil  \pm \totzaadsysnofegmultil$   &
$\zaanofmugmultil \pm \totzaadsysnofmugmultil$  &
$\zaanoflgmultil  \pm \totzaadsysnoflgmultil$   \\
\hline
$\Z+$Jet, jet faking $\gamma$   & 
$\nofejgmultil  \pm \totdsysnofejgmultil$       &
$\nofmujgmultil \pm \totdsysnofmujgmultil$      &
$\nofljgmultil  \pm \totdsysnofljgmultil$       \\
$\Z\goes \epem, e\goes\gamma$           & 
$\efpnofegmultil \pm \totefpdsysnofegmultil$    &
--                                       &
$\efpnofegmultil \pm \totefpdsysnofegmultil$    \\
QCD (Non-WZ) fakes                         & 
$\jqcdnofegmultil  \pm \totjqcddsysnofegmultil $   &
$\jqcdnofmugmultil \pm \totjqcddsysnofmugmultil$   &
$\jqcdnoflgmultil  \pm \totjqcddsysnoflgmultil $   \\
DIF (Decays-In-Flight)                     & 
--                                                &
$\difnofmugmultil \pm \totdifdsysnofmugmultil$   &
$\difnoflgmultil  \pm \totdifdsysnoflgmultil $   \\
\hline
Total                                              & 
$\smnofegmultil  \pm \totdsysnofegmultil (tot)$    &
$\smnofmugmultil \pm \totdsysnofmugmultil(tot)$    &
$\smnoflgmultil  \pm \totdsysnoflgmultil (tot)$    \\
\hline
Observed                                 & 
$\nofegmultil$                           & 
$\nofmugmultil$                          &                 
$\noflgmultil$                           \\
\hline
\end{tabular}

\label{llg.table}
\end{center}
\end{table}

The predicted and observed totals for the $\lgmet$ and $\llg$ searches
are shown in Tables~\ref{lgmet.table} and~\ref{llg.table},
respectively. We observe $\noflglgmet$ $\lgmet$ events, versus the
expectation of $\smnoflglgmet \pm \totdsysnoflglgmet$ events. The
predicted and observed kinematic distributions for the $\lgmet$
signature (the sum of electrons and muons) are compared in
Figures~\ref{wg_fig1_leptons} and~\ref{wg_fig2_leptons}~\cite{no_overflows}.

In the $\llg$ channel, we observe $\noflgmultil$ events, versus an
expectation of $\smnoflgmultil \pm \totdsysnoflgmultil$ events. There
is no significant excess in either signature. The predicted and
observed kinematic distributions for the $\llg$ signature are compared
in Figures~\ref{zg_fig1_leptons},~\ref{zg_fig2_leptons}
and~\ref{zg_fig3_leptons}~\cite{no_overflows}.

The predicted and observed totals for the $\lgg$ and $e\mug$ searches
are shown in Tables~\ref{emug.table} and~\ref{lgg.table},
respectively. We observe no $\lgg$ or $e\mug$ events, versus the
expectation of $\smnoflgmultig \pm \totdsysnoflgmultig$ and
$\smnoflgemug \pm \totdsysnoflgemug$ events, respectively.

\begin{table}[!t]
\begin{center}
\caption{A comparison of the numbers of events predicted by the
SM and the observations for the $e\mu\gamma$ signature. The SM
predictions are dominated by $\Zg$
production~\cite{MadGraph,Baur,CompHep}. Other contributions come from
$\Wg$, $\Zgg$, $\Wgg$, and misidentified leptons, photons, or $\met$.}
\begin{tabular}{|l|c|}
\hline
\multicolumn{2}{|c|} {
{\bf CDF Run II Preliminary, $\apprluminosity\invpb$}}\\[3pt]
\hline
\multicolumn{2}{|c|}{
$e\mu$ + Photon + X Events}\\
\hline
Standard Model Source & $e\mu\gamma+X$  \\
\hline
$\Z\gamma$            		          & 
$\zanoflgemp  \pm \zadnoflgemp$     \\
$W\gamma$            		          & 
$\wanoflgemp  \pm \wadnoflgemp$     \\
$\Z\gamma\gamma$ 	     	& 
$\zaanoflgemp  \pm \zaadnoflgemp$   \\
$W\gamma\gamma$ 	     	& 
$\waanoflgemp  \pm \waadnoflgemp$   \\
\hline
$e\mu j$, $j\rightarrow\gamma$   & 
$\nofljgemp  \pm \dnofljgemp$       \\
$ee\mu$, $e\goes\gamma$                & 
$\efpnofegemp  \pm \efpdnofegemp$   \\
\hline
$\tau\gamma$ contribution               & 
$\wazataunoflgemp  \pm \wazataudnoflgemp $  \\
\hline
Total & $\smnoflgemug \pm \totdsysnoflgemug$    \\
\hline
Observed                         & 
$\noflgemp$                              \\
\hline
\end{tabular}

\label{emug.table}
\end{center}
\end{table}

\begin{table}[!t]
\begin{center}
\caption{A comparison of the numbers of events predicted by the
SM and the observations for the $\lgg$ signature.
The SM predictions are
dominated by $\Zgg$ production~\cite{MadGraph,CompHep}. Dominant
contribution comes from misidentified photons.}
\begin{tabular}{|l|c|c||c|}
\hline
\multicolumn{4}{|c|} {
{\bf CDF Run II Preliminary, $\apprluminosity\invpb$}}\\[3pt]
\hline
\multicolumn{4}{|c|}{
Multi-Photon + Lepton Events} \\
\hline
SM Source & $\egg$ & $\mugg$ & $\lgg$ \\
\hline
$W^{\pm}\gamma\gamma$                     & 
$\waanofegmultig  \pm \waadnofegmultig$   &
$\waanofmugmultig \pm \waadnofmugmultig$  &
$\waanoflgmultig  \pm \waadnoflgmultig$   \\
$\Z\gamma\gamma$                        & 
$\zaanofegmultig  \pm \zaadnofegmultig$   &
$\zaanofmugmultig \pm \zaadnofmugmultig$  &
$\zaanoflgmultig  \pm \zaadnoflgmultig$   \\
$\Z\gamma$, $e\goes\gamma$                & 
$\efpnofegmultig  \pm \efpdnofegmultig$   &
-                                         &
$\efpnofegmultig  \pm \efpdnofegmultig$   \\
$\ell jj$, $\ell\gamma j$, $j\goes\gamma$ & 
$\nofejgmultig  \pm \totdsysnofejgmultig$       &
$\nofmujgmultig \pm \totdsysnofmujgmultig$      &
$\nofljgmultig  \pm \totdsysnofljgmultig$       \\
\hline
Total             & 
$\smnofegmultig  \pm \totdsysnofegmultig $    &
$\smnofmugmultig \pm \totdsysnofmugmultig$   &
$\smnoflgmultig  \pm \totdsysnoflgmultig $    \\
\hline
Observed                                 & 
$\nofegmultig$                           & 
$\nofmugmultig$                          & 
$\noflgmultig$                           \\
\hline
\end{tabular}

\label{lgg.table}
\end{center}
\end{table}

\section{Conclusions}
\label{conclusions}


To test whether something new was really there in either the
$\llggmet$ or $\lgmet$ signatures in Run~I, we have repeated the
$\lgX$ search with the same kinematic requirements as the Run I
search, but with an exposure more than 10 times larger,
$\lumi\pm\dlumi$ $\invpb$, a higher $\ppbar$ collision energy, 1.96
$\TeV$, and the CDF II detector~\cite{CDFII}.

We observe $\noflglgmet$ $\lgmet$ events, versus an expectation of
$\smnoflglgmet \pm \totdsysnoflglgmet$ events from SM physics and
background sources. In the $\llg$ channel, we observe $\noflgmultil$
events, versus an expectation of $\smnoflgmultil \pm
\totdsysnoflgmultil$ events. There is no significant excess in either
signature. We can conclude that the 2.7 $\sigma$ effect observed in Run I,
measured with the same criteria and a very similar detector, was a
fluctuation. 

With respect to the Run I $\eeggmet$ event, we observe no $\lgg$
events versus an expectation of $\smnoflgmultig \pm
\totdsysnoflgmultig$ events. We do find 3 $\llg$ events with $\met>
25$ GeV, versus an expectation of $0.6 \pm 0.1$ events, corresponding
to a likelihood of 2.4 \%. We do not consider this significant, and
there is nothing in these 3 events to indicate they are due to
anything other than a fluctuation.  The $\eeggmet$ event thus remains
a single event selected {\it a posteriori} as interesting, but whether
it was from SM $WW\gamma\gamma$ production, a rare background, or a
new physics process we cannot determine.

Lastly, we observe no  $e\mug$ events, versus a SM expectation of 
$\smnoflgemug \pm \totdsysnoflgemug$ events.


%



\begin{thebibliography}{99}

\bibitem{EtPt} Transverse momentum and energy are defined as $\Pt =
p\sin\theta$ and $\Et = E\sin\theta$, respectively.  
%
Missing $\rm E_T$ ($\metvec$) is defined by $\metvec = -\sum_{i} E_T^i
\hat{n}_i$, where i is the calorimeter tower number for $|\eta| <$ 3.6
(see Ref.~\cite{CDF_coo}), and $\hat{n}_i$ is a unit vector
perpendicular to the beam axis and pointing at the i$^{th}$
tower. We correct $\metvec$ for jets and muons. We define
the magnitude $\met=|\metvec|$.
%
We use the convention that ``momentum'' refers to $pc$ and ``mass'' to $mc^2$.

\bibitem{SM} S.L.~Glashow,
Nucl. Phys. {\bf 22} 588, (1961); S. Weinberg,
Phys. Rev. Lett. {\bf 19} 1264, (1967);
A. Salam, Proc. 8th Nobel Symposium, Stockholm, (1979).

\bibitem{Toback_all}
F.~Abe \textit{et al.} (CDF Collaboration), Phys. Rev. D \textbf{59},
092002 (1999); F.~Abe \textit{et al.} (CDF Collaboration),
Phys. Rev. Lett. \textbf{81}, 1791 (1998); D.~Toback, Ph.D. thesis,
University of Chicago, 1997.

\bibitem{Jeff_all} 
D. Acosta \textit{et al.} (CDF Collaboration), Phys. Rev. D
\textbf{66}, 012004 (2002); 
D. Acosta \textit{et al.} (CDF Collaboration), Phys. Rev. Lett. \textbf{89},
041802 (2002); 
J.~Berryhill, Ph.D. thesis, University of Chicago, 2000.

\bibitem{Loginov_all} 
A.~Abulencia \textit{et al.} (CDF Collaboration), Phys. Rev. Lett. \textbf{97},
031801 (2006); 
A.~Loginov (for the CDF Collaboration), Eur.Phys.J. C \textbf{46},
Supplement 2, pp. 21-31 (2006); 
A. Loginov, Ph.D thesis, Institute for
Theoretical and Experimental Physics, Moscow, Russia, September,
2006.

\bibitem{CDFI} 
F. Abe \textit{et al.} (CDF Collaboration),
Nucl. Instrum. Methods A \textbf{271}, 387 (1988).

\bibitem{COT} 
A. Affolder \textit{et al.}, 
Nucl. Instrum. Methods A \textbf{526}, 249 (2004).

\bibitem{SVX} 
A. Sill \textit{et al.}, 
Nucl. Instrum. Methods A \textbf{447}, 1 (2000);
A. Affolder \textit{et al.},
Nucl. Instrum. Methods A \textbf{453}, 84 (2000); 
C.S. Hill,
Nucl. Instrum. Methods A \textbf{530}, 1 (2000).

\bibitem{CDF_coo} The CDF coordinate system of 
$r$, $\varphi$, and $z$ is cylindrical, with the $z$-axis along the
proton beam. The pseudorapidity is $\eta = -\ln(\tan(\theta/2))$.

\bibitem{cal_upgrade} S. Kuhlmann \textit{et al.}, 
Nucl. Instrum. Methods A \textbf{518}, 39, 2004.

\bibitem{muon_systems} 
The CMU system consists of 
gas proportional
chambers in the region $|\eta|<0.6$; the CMP system consists of
chambers after an additional meter of steel, also for
$|\eta|<0.6$. The CMX chambers cover 
$0.6<|\eta|<1.0$.

\bibitem{jet_corr} 
A. Bhatti \textit{et al.}, accepted to 
Nucl. Instrum. Methods, 2006. 

\bibitem{no_overflows} There are no overflows in any of the figures in
  this paper.
      
\bibitem{susy} 
S.~Ambrosanio, G.L.~Kane, G.D.~Kribs, S.P.~Martin, and S.~Mrenna,
Phys. Rev. D \textbf{55}, 1372 (1997); B.C.~Allanach, S.~Lola,
K.~Sridhar, Phys. Rev. Lett. \textbf{89}, 011801 (2002).

\bibitem{CDF_WZgamma} 
D. Acosta \textit{et al.} (CDF Collaboration), Phys. Rev. Lett. \textbf{94},
041803 (2005).

\bibitem{MadGraph} 
T. Stelzer and W. F. Long, Comput. Phys. Commun. \textbf{81}, 357
(1994); F. Maltoni and T. Stelzer, JHEP \textbf{302}, 27 (2003).

\bibitem{Baur} 
U. Baur, T. Han, and J. Ohnemus,
Phys. Rev. D \textbf{48}, 5140 (1993);
J. Ohnemus, Phys. Rev. D \textbf{47}, 940 (1993).

\bibitem{CompHep} 
E. Boos \textit{et al.} (The {\sc comphep} Collaboration),
Nucl. Instrum. Methods A \textbf{534}, 250, (2004). 

\bibitem{Pythia} 
T.~Sjostrand, Comput. Phys. Commun. \textbf{82} (1994) 74;
S.~Mrenna, Comput. Phys. Commun. \textbf{101} (1997) 232.
 
\bibitem{Baur_NLO} U.~Baur, T.~Han and J.~Ohnemus, 
Phys.\ Rev.\ D {\bf 48}, 5140 (1993);
U.~Baur, T.~Han and J.~Ohnemus, 
Phys.\ Rev.\ D {\bf 57}, 2823 (1998). 
Both the $\Wg$ and $Z\gamma$ K-factors are fixed at 1.36 for generated
$\ell\nu$ masses below 76 $\GeV$ and for generated $\lplm$ masses below
86 $\GeV$. Above the poles the K-factors grow with $\Et^{\gamma}$ to
be 1.62 and 1.53 at $\Et^{\gamma}=100$ $\GeV$ for $\Wg$ and $Z\gamma$,
respectively.

\bibitem{QCD_background} In each signature the background distribution
is derived from the observed distribution; the background level can
thus be seen to follow the data in the appropriate figures. The
advantage of this procedure (as opposed to just cutting on the track
isolation variable) for the low statistics on the tails of the
distribution is that one can get some sense of the level of background
from rare fragmentations of jets that may be topology dependent.

\bibitem{decay_in_flight} Decays before or after the COT volume result in a
correct measurement of the momentum and are included in the other
background estimate.

\bibitem{CDFII} 
D. Acosta \textit{et al.} (CDF Collaboration), Phys. Rev. D \textbf{71}, 032001
(2005).

\end{thebibliography}
\end{document}